\documentclass[showpacs,amsmath,amssymb,pre,superscriptaddress,floatfix,twocolumn]{revtex4-1}
\usepackage{graphicx}% Include figure files
\usepackage{float}
\usepackage{dcolumn}% Align table columns on decimal point
\usepackage{bm}% bold math
\usepackage{color}
\usepackage[toc,page]{appendix}

\usepackage[english]{babel}
\usepackage[utf8]{inputenc}
\newcommand{\lt}{\left(}
\newcommand{\rt}{\right)}
\newcommand{\lqq}{\left[}
\newcommand{\rqq}{\right]}
\newcommand{\lan}{\left\langle}
\newcommand{\ran}{\right\rangle}

\newcommand{\av}[1]{\lan #1 \ran}

\newcommand{\ket}[1]{\left| #1 \ran}
\newcommand{\bra}[1]{\lan #1 \right|}

\newcommand{\comm}[2]{\left[ #1, #2 \right]}

\newcommand{\be}{\begin{equation}}
\newcommand{\ee}{\end{equation}}

\newcommand{\suml}[2]{\sum\limits_{#1}^{#2}}

\newcommand{\nn}{z}

\newcommand{\matb}{\left(\begin{array}}
\newcommand{\mate}{\end{array}\right)}
\newcommand{\nol}{\nonumber\\}
\newcommand{\rmd}{{\rm{d}}}

\newcommand{\ha}{\frac{1}{2}}

\usepackage{xr}
\usepackage{xcite}

\externalcitedocument[S-]{Supplemental5}% <- full or relative path

\begin{document}

\title{Absorbing state phase transition with competing quantum and classical fluctuations}

\author{Matteo Marcuzzi}
\affiliation{School of Physics and Astronomy, University of Nottingham, Nottingham, NG7 2RD, United Kingdom}
\author{Michael Buchhold}
\affiliation{Institut f\"ur Theoretische Physik, Universit\"at zu K\"oln, D-50937 Cologne, Germany}
\author{Sebastian Diehl}
\affiliation{Institut f\"ur Theoretische Physik, Universit\"at zu K\"oln, D-50937 Cologne, Germany}
\author{Igor Lesanovsky}
\affiliation{School of Physics and Astronomy, University of Nottingham, Nottingham, NG7 2RD, United Kingdom}

\date{\today}% It is always \today, today,

\begin{abstract}
Stochastic processes with absorbing states feature remarkable examples of non-equilibrium universal phenomena. While a broad understanding has been progressively established in the classical regime, relatively little
is known about the behavior of these non-equilibrium systems in the presence of quantum fluctuations. 
Here we theoretically address such a scenario in an open quantum spin model which in its classical limit undergoes a directed percolation phase transition. By mapping the problem to a non-equilibrium field theory, we show that the introduction of quantum fluctuations stemming from coherent, rather than statistical, spin-flips alters the nature of the transition such that it becomes first-order. In the intermediate regime, where classical and quantum dynamics compete on equal terms, we highlight the presence of a  bicritical point with universal features different from the directed percolation class in low dimension. We finally propose how this physics could be explored within gases of interacting atoms excited to Rydberg states.
\end{abstract}

\maketitle
\section{Introduction} Non-equilibrium phenomena are ubiquitous in nature, ranging from the microscopic scales of chemical reactions to the macroscopic ones of disease-spreading. Remarkably, analogously to the equilibrium case, non-equilibrium ensembles can show the emergence of universal behavior, signaling the irrelevance of the microscopic details of the dynamics for macroscopic observables. This occurs when such out-of-equilibrium systems start to act collectively  \cite{Ma, QPT, Tauber-book, NEQ_PT1}. On a fundamental level, a distinction arises depending on the presence or absence of detailed balance \cite{Microrev1, Microrev2, Sieberer15, Eq_vs_NonEq}, between systems which evolve towards a stationary equilibrium state (e.g., quenched systems coupled to thermal baths \cite{Janssen}) or that preserve their non-equilibrium character even in the long-time limit, representing flux equilibrium states. The universal dynamical features of purely \emph{classical} systems have been extensively studied and classified both for unbroken \cite{HH} and broken \cite{DP_Hinrichsen, NEQ_lattice, Racz2002, KPZ} detailed balance, i.e., genuine non-equilibrium systems.
Recently, experiments in various platforms have started to systematically probe driven open \emph{quantum} systems. The spectrum includes light-driven semiconductor heterostructures \cite{Carusotto13}, arrays of driven microcavities \cite{Hartmann2008,Tomadin2010}, 
%
%
%laser-driven vapour cells \cite{Carr2013, Urvoy2015, Sibalic2015},  
cold atoms in optical lattices \cite{Schaub2015}, cavities \cite{Ritsch2013, Brennecke2012} and microtraps \cite{Nogrette2014, Schlosser2011, Dumke2002}. Several among these instances employ excitation of the atoms to high-lying Rydberg orbitals \cite{Gallagher84, Ryd-QI, Rydberg2} in order to achieve strong interatomic interactions and to study cooperative effects \cite{Carr2013, Valado2015, Urvoy2015}.
In all these systems, the driving/dissipation introduces coherence loss and explicitly violates the equilibrium conditions at the microscopic level \cite{Sieberer15,SiebererRev}.
It is thus a challenge to identify to what extent the non-equilibrium and the quantum nature of the dynamics impact on the macroscopic phase diagram and phase transition properties. Oftentimes, upon coarse graining such systems lose their quantum character and equilibrium conditions are effectively restored \cite{mitra06,dallatorre10,Dalla13,Sieberer2013,Th-Mitra1,Th-Mitra2}. But there are instances where non-equilibrium \cite{Altman15, Prosen2010} and quantum \cite{Lee13,Marino2015} aspects persist even at asymptotically large wavelength.

\emph{Directed percolation} (DP) represents an instance of a classical, but intrinsically non-equilibrium system (for a review, see \cite{DP_Hinrichsen}). Despite its robustness, its experimental observation has so far been elusive \cite{Hinrichsen_exp}, with a single remarkable exception \cite{DP_exp, DP_explong}. It was suggested recently to realize and explore DP dynamics in 
%the long wavelength properties of 
cold gases of atoms excited to high-lying Rydberg states \cite{Marcuzzi2015}. In that case, the non-equilibrium nature of dynamics persists macroscopically, but the impact of quantum dynamics fades out completely under coarse graining. 

In this work, we harness the opportunities that result from the fact that such Rydberg gases indeed represent driven open \emph{quantum} systems to go beyond the realm of classical physics, and establish a novel absorbing state phase transition characterized by the interplay of classical and quantum terms on equal footing. This transition does not fall into the DP universality class, and its origin can be unambiguously traced back to the presence of coherent dynamics. More precisely, the latter introduces a strong coupling first-order non-equilibrium phase transition without counterpart in the purely classical DP problem. Remarkably, this discontinuous phase transition terminates in a novel bicritical point which even asymptotically at large distances and in dimensions $d<2$, does not feature the symmetries underlying DP, or any equilibrium problem. %We also suggest how to implement this physics in experiments with cold atomic Rydberg gases.

\section{Model}
In the following we reproduce a quantum variant of the \emph{contact process} (for an introduction we refer to Ref.~\cite{DP_Hinrichsen}). Its defining property is the following: in a lattice of ``active'' and ``inactive'' sites, the former can spontaneously decay to inactive, whereas activation can only occur in the proximity of already active sites. Thus, the fully-inactive state is absorbing, i.e., once reached it cannot be left. 
Specifically, we consider a lattice of quantum two-level systems with spacing $r$. On every site $k$ we define the basis $\ket{a_k}$ (active) and $\ket{i_k}$ (inactive), the density of active sites $n_k = \ket{a_k} \bra{a_k}$ and the ladder operators $\sigma_k^+ = \ket{a_k} \bra{i_k}$ and $ \sigma_k^- = \ket{i_k} \bra{a_k}$.  
%
%%%%%%%%
%%%%%%%%%%
%%%%%%%%%
%
\begin{figure}[ht!]
%\begin{center}
%\centering
	\includegraphics[width=0.8\columnwidth]{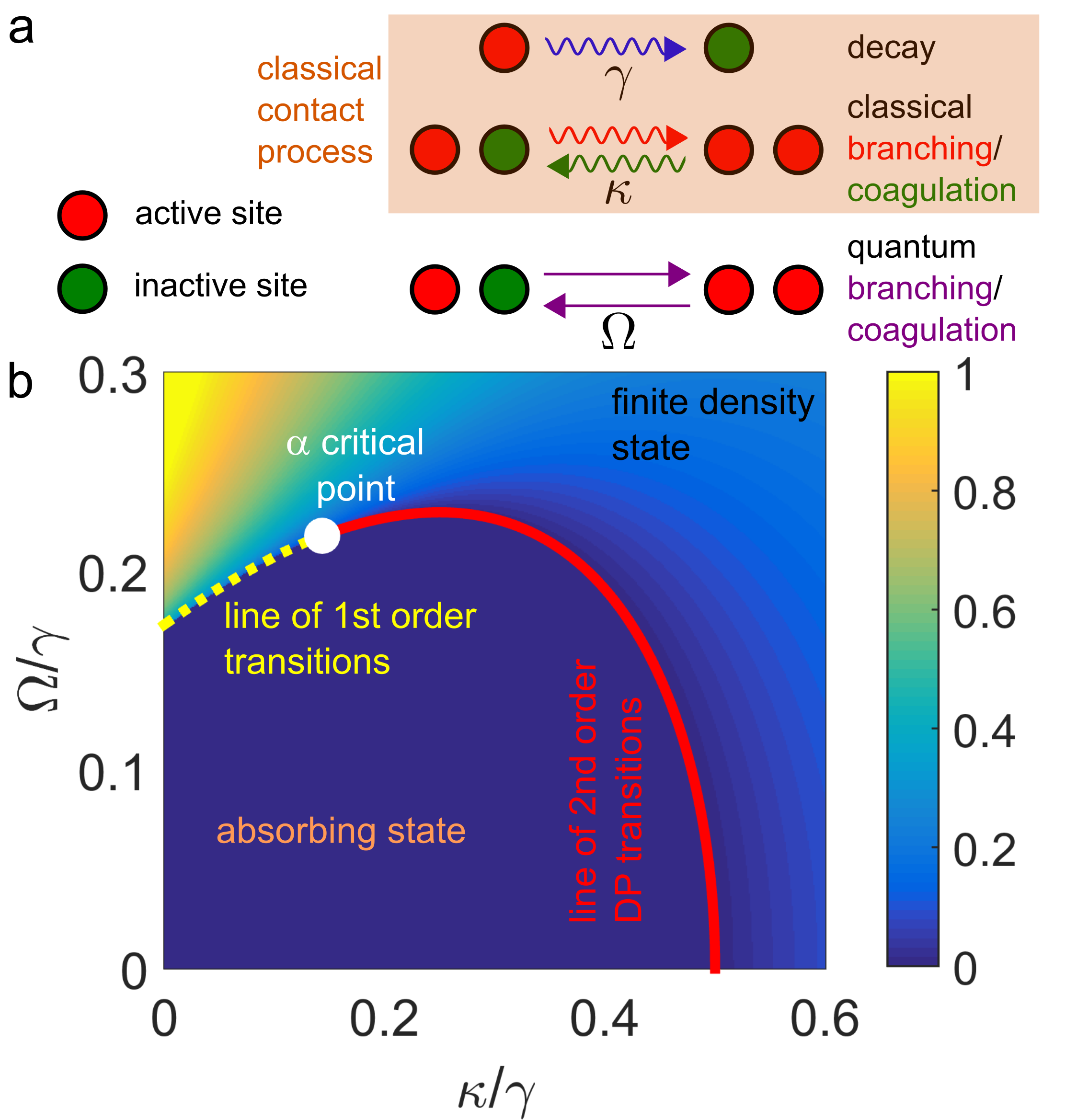}
%\end{center}
\caption{(Color online) (a) Fundamental processes. We consider a lattice whose sites admit two states: active (red) and inactive (green). Active sites decay at a rate $\gamma$ and become inactive. Proliferation of active sites is possible through classical (rate $\kappa$) and quantum (strength $\Omega$) branching. 
%The time-reversed processes (coagulation) are also present. 
(b) Phase diagram constructed from the effective action \eqref{eq:action} in saddle-point approximation (color code corresponds to density of active sites). In the classical limit ($\Omega=0$) the system exhibits a continuous (2nd order) directed percolation phase transition between an absorbing state and one with finite density. This transition extends into the quantum regime (thick red line) until the critical point $\alpha$ is reached. In the quantum limit ($\kappa=0$) a first-order transition is found which also extends into the classical regime (dashed yellow line) up to point $\alpha$. In the neighborhood of this line, a narrow region of coexistence of two attractive stationary solutions is present, which is not resolved here. The high values of the density reached in the active phase stem from neglecting higher orders in $n$ in the action, which would otherwise prevent it from exceeding $1/2$.  
}
\label{fig:1}
\end{figure}
Under the action of Markovian noise sources, the state $\rho$ of the system evolves according to the Lindblad equation \cite{Lindblad76, Breuer_P} ${\dot{\rho} = -i \comm{H}{ \rho} + \suml{a,k}{} \mathbf{D}[L_{a,k}]\rho}$ [see sketch in Fig.~\ref{fig:1}], where
\be
	H = \Omega \sum_k \Pi_k\,\sigma_k^x   \quad \text{with} \quad \Pi_k = \sum_{j\, \text{nn} \, k} n_j
	\label{eq:H}
\ee
is the quantum Hamiltonian, $\sigma_k^x = \sigma_k^+ + \sigma_k^-$, and ``$ \text{nn} \, k$'' denotes nearest neighbors (nn) of site $k$; ${\mathbf{D}[X]\rho=X \rho X^\dagger - (X^\dagger X\rho + \rho X^\dagger X)/2}$ is the dissipator and $L_{a,k}$ are the so-called jump operators, having the index $a$ labeling the process type, and $k$ the lattice site. These jump operators are chosen to define a modified contact process \cite{DP_Hinrichsen}, which is known to feature a DP transition, and include \emph{decay} $L_{\text{d},k} = \sqrt{\gamma} \sigma_{k}^-$ ($\ket{a_k} \to \ket{i_k}$) and --- for every neighbor $j$ of $ k$ ---  \emph{branching} $L_{\text{b},j,k} = \sqrt{\kappa} \, n_j \, \sigma_k^{+}$ (an active site can activate a neighboring one $\ket{a_j i_k} \to \ket{a_j a_k}$) and \emph{coagulation} $L_{\text{c},j,k} = \sqrt{\kappa} \, n_j \, \sigma_k^{-}$ (the inverse process $\ket{a_j a_k} \to \ket{a_j i_k}$). The operator $\Pi_k $ in $H$ represents the simplest choice which effectively reproduces the requirement of an active site nearby to flip a spin; this makes $H$ the ``minimal quantum equivalent'' of the noisy branching/coagulation above. Similar ``constrained'' Hamiltonians have been studied in the past with a focus on many-body localization \cite{vanHorssen2015, Hickey2014}.
%
%%%%%%%%
%%%%%%%%%%
%%%%%%%%%

%

\section{Equations of motion and density path integral}
We infer here the properties of the phase diagram by exploiting an
%In order to more clearly understand the phase diagram structure emerging from this problem, we reformulate it with an
effective path integral description for the density variable $n_k$ alone.
We start by deriving the Heisenberg-Langevin equations of motion (EOM) \cite{Scully} for the single-site operators $n_k$, $\sigma^x_k$ and $\sigma^y_k= -i\sigma^+_k  + i\sigma^-_k$. For convenience we introduce the coordination number $z$ (number of nearest neighbors per lattice site), the shorthand $\Sigma^{x/y}_k = \sigma^{x/y}_k\sum_{j \, \text{nn}\, k} \sigma^{x}_j$, rescale time by $t \to \tau = \gamma t$ and the rates accordingly, i.e. $\chi=\kappa/\gamma $ and $ \omega=\Omega/\gamma$:
\begin{align}
\dot n_k&=-n_k+\left[\omega \sigma^y_k-\chi(2n_k-1)\right]  \Pi_k +\hat{\xi}^n_k,\label{eq:quantum_n}\\[1mm]
\dot\sigma^x_k&=\omega \Sigma_k^y - \tfrac{z\chi + 1}{2}\sigma^x_k - \chi\sigma^x_k  \Pi_k +\hat{\xi}^x_k,\label{eq:quantum_x}\\[1mm]
\dot\sigma^y_k&=\omega \Sigma^x_k-\tfrac{z\chi + 1}{2}\sigma^y_k -   
%+ \nonumber \\   &-& 
  \left[\omega(4n_k-2)
+\chi\sigma^y_k\right] \Pi_k + \hat{\xi}^y_k. \label{eq:quantum_y}
\end{align}
The quantum noise terms $\hat{\xi}^{\alpha}_k$ consider the fluctuations of the bath and depend on the structure of the jump operators. They show vanishing averages but non-trivial, Markovian correlations, which for the present setup are (in rescaled units)
$\langle \hat{\xi}^x_k\hat{\xi}^x_{k'}\rangle=\langle \hat{\xi}^y_k\hat{\xi}^y_{k'}\rangle=\delta_{k,k'}$, $\langle \hat{\xi}^n_k\hat{\xi}^n_{k'}\rangle=\delta_{k,k'} n_k$, $\langle  \hat{\xi}^x_k  \hat{\xi}^{y}_{k'}   \rangle = -i\delta_{k,k'}$, $\langle  \hat{\xi}^n_k  \hat{\xi}^{x}_{k'}   \rangle = -\delta_{k,k'} \sigma_k^{+}$ and $\langle  \hat{\xi}^n_k  \hat{\xi}^{y}_{k'}   \rangle = i\delta_{k,k'} \sigma_k^{+}$
up to leading order in the density (see Appendix \ref{app:HL_eqs}).

In the following,  we work in the continuum limit $(k,t) \to (\vec{x}, t) \equiv X$ and derive an effective path integral for the density field $n_X$ via a Martin-Siggia-Rose (MSR) construction \cite{MSR, MSR+J, MSR+D, Tauber-book}, presented in Appendix \ref{app:path_int}. Crucially, the $\sigma^{x,y}$-fields are gapped, and thus can be integrated out perturbatively. The resulting long wavelength field theory depends on the density variable $n$ alone, and is obtained by additionally performing a derivative expansion of the action.
It reads
\begin{align}
	S_n&=\int_{X} \tilde{n}_X \Big[(\partial_t - D \nabla^2  +\Delta)n_{X}+ u_3n^2_{X} + u_4n_{X}^3\Big] \nonumber\\
& - \int_{X} \left[\tfrac{1}{2} \tilde{n}_X^2 n_X + \mu_4 \tilde{n}_X^2 n_X^2   \right]  \equiv S^{(1)}_n + S^{(2)}_n,
\label{eq:action}
\end{align}
where $D = r^2 \chi$ represents a diffusion constant (lattice spacing $r$) and ${\Delta= 1 - z\chi - \tfrac{8z^2\omega^2}{(z\chi+1)^3}}$, $u_3=2z \left(\chi- \tfrac{2z\omega^2}{z\chi+1}\right)$, $u_4=\tfrac{ 8z^2 \omega^2}{z\chi+1}$ and $\mu_4 =  \tfrac{2z^2 \omega^2}{(z\chi + 1)^2} +   \tfrac{128 z^4 \omega^4}{(z\chi + 1)^6}$ are the microscopic coupling constants. The \emph{response field} $\tilde{n}$ encodes the linear response properties of $n$ under small perturbations.

At this point we emphasize two key properties of the action \eqref{eq:action}: First, the absence of a density independent Markovian noise level $\sim T\tilde{n}_X^2$ (which is necessarily present in any classical system in thermal equilibrium). This is characteristic of DP dynamics, which feature the absence of density fluctuations in the absorbing state $n_X=0$ and consequently a multiplicative kernel $\sim n_X$. An additive noise introduced by the dissipative terms $L_d=\sqrt{\gamma}\sigma^-$ only occurs in the eliminated spin variables $\sigma^{x,y}$.
Second, the presence of a non-zero coherent coupling $\omega\neq 0$ -- i.e. the intrinsic quantum effect -- leads to the appearance of non-zero couplings $u_4$ and $\mu_4$ as well as a negative contribution to $u_3$. This new ``quantum'' scale $\omega$ breaks a fundamental symmetry of the DP class (specified below) and strongly modifies the phase diagram compared to the purely dissipative model [see Fig. \ref{fig:1}].

\section{Effective potential and mean-field phase diagram}
The discussion of the various phases and phase transitions of the system is considerably simplified by realizing that the deterministic contribution to the action $S_n^{(1)}$ can be written as $ \int_{X} \tilde{n}_{X}\Big[\partial_tn_{X}- D \nabla^2n_{X}+\frac{\delta\Gamma(n_{X})}{\delta n_{X}}\Big]$, where
\begin{eqnarray}
\Gamma(n)&=&\frac{\Delta}{2}n^2+\frac{u_3}{3}n^3  +  \frac{u_4}{4}n^4 \label{Eq10}
\end{eqnarray}
is a local effective potential. In the absence of fluctuations $\Gamma$ characterizes the mean-field phases, which are determined by the properties around its minima.

The corresponding phase diagram is shown in  Fig.~\ref{fig:1}(b). 
The active phase is identified by $\Delta < 0$, $u_4 \geq 0 $ and $u_3>0$, which leads to a single minimum of the effective potential at finite density. On the other hand, when both $\Delta$ and $u_4$ are positive, there is a local minimum of $\Gamma$ at $n=0$. For negative and sufficiently strong cubic coupling $u_3 < -2\sqrt{u_4\Delta}$, there exists a second local minimum at finite density $n>0$. In this regime, the mean field evolution features two attractive fixed points and the thermodynamic phase is determined within the optimal path approximation in phase space~\cite{Kamenev}.

\begin{figure}[ht!]
	\includegraphics[width=1.0\columnwidth]{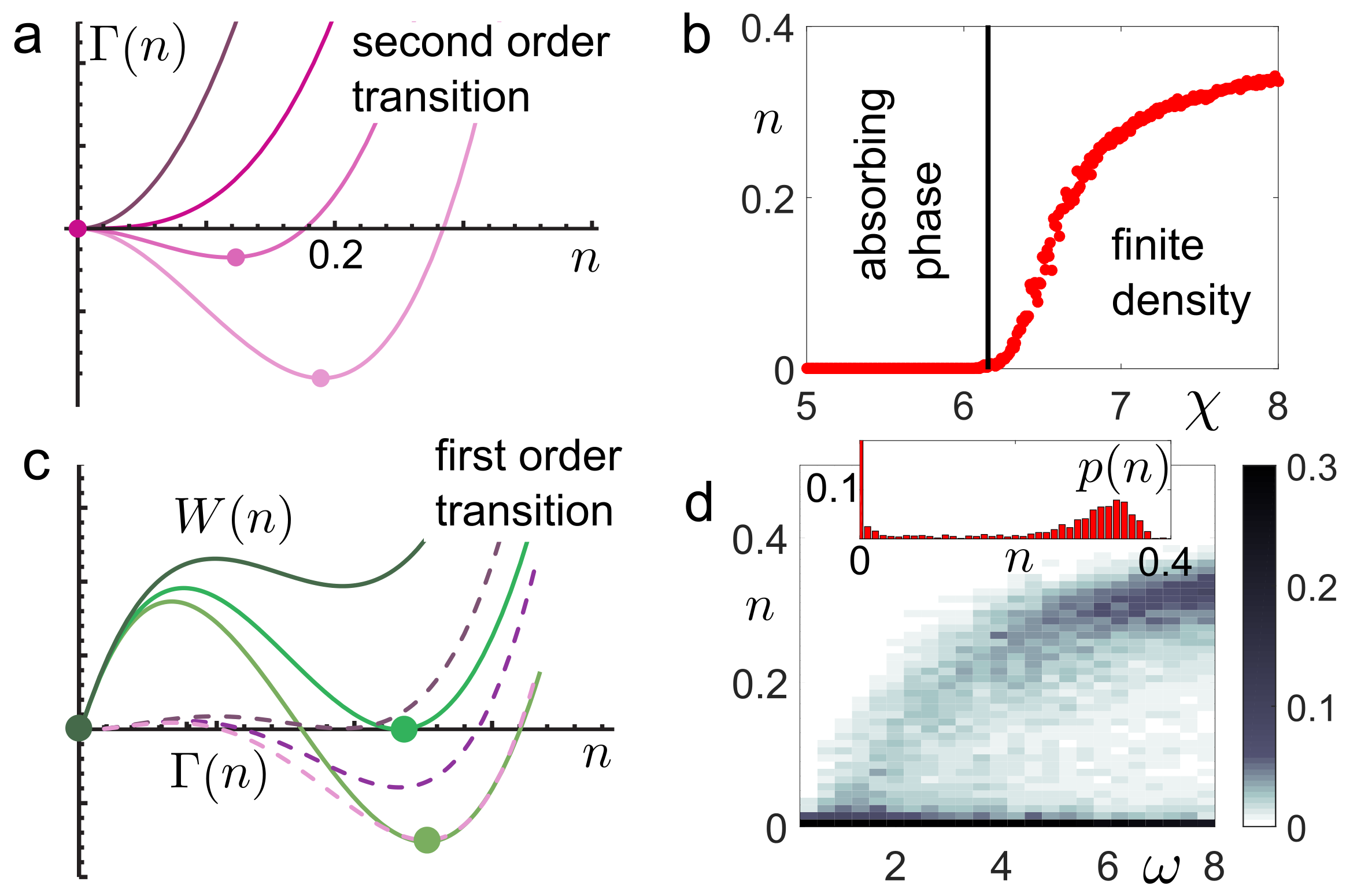}
\caption{
Effective potential and phase transitions. (a) Behavior of the effective potential $\Gamma(n)$ (arbitrary units) across the second order phase transition. Dots mark the minima of $\Gamma(n)$. The transition occurs when $\Delta$ in Eq.~(\ref{Eq10}) changes sign. (b) Stationary state density in the classical limit ($\omega=0$) as a function of $\chi$ (chain of $200$ sites, average over $10^3$ realizations per point), obtained via Monte Carlo simulations starting from a completely active configuration and stopped at time $\gamma t = 10^4$. 
The data show the characteristic behavior of a second order phase transition around $\chi_c\approx 6.2$. (c) Effective potential $\Gamma(n)$ (dashed lines) and corresponding ``optimal-path'' potential $W(n)$ (solid lines), see Eq.~\eqref{opath}, across the first-order transition. 
At the transition point, $W(n_1 = 0) = W(n_2) = 0$. (d) Steady-state histogram of the density in the quantum limit $\chi=0$ ($12$ spins) obtained via a quantum-jump Monte Carlo (QJMC) method,  indicating a first-order transition ($\omega_c\approx 2$) as $\omega$ increases. Two stable stationary solutions, one with zero and one with finite density, emerge. The inset displays a section of the histogram taken at $\omega=8$.  }
\label{fig:2}
\end{figure}

Three different types of phase transitions from the active to the inactive state can be thus identified, their nature depending on the specific choice of parameters and the dimensionality. 
When the gap $\Delta$ vanishes with both $u_3,u_4>0$ the system undergoes a second order phase transition [see Fig.~\ref{fig:2}(a)], corresponding to a diverging correlation length $\xi=1/\sqrt{|\Delta|}\rightarrow\infty$.
Numerical evidence for this transition is presented in panel (b) of Fig.~\ref{fig:2}, which displays the stationary density of active sites obtained for $\omega =0$ in a chain of $200$ sites.
For $\Delta > 0$ and $u_3 \leq -2\sqrt{u_4 \Delta}$, the transition from the active to the inactive phase takes place instead at finite correlation length $\xi=1/\sqrt{|\Delta|}<\infty$. The form of the effective potential $\Gamma(n)$ suggests a first-order transition line in this regime featuring the coexistence of the zero and finite-density solutions. 
This case, however, requires additional care due to the specific form of the noise, as detailed further below.

The $\alpha$ point in Fig.~\ref{fig:1}(b) located at $\Delta = u_3 = 0$ represents a bicritical point at which both the line $(\Delta>0, u_3=-2\sqrt{\Delta u_4})$ and the line of second order transitions $(\Delta = 0, u_3>0)$ terminate. At this point, the quartic potential term $u_4$ provides the leading non-linearity.

\section{Fluctuations at the continuous transition}
The competition between quantum and classical dynamics strongly affects the nature of the active-to-inactive transition. In the absence of the coherent coupling, $u_4, \mu_4=0$, the action \eqref{eq:action} is equivalent to the so-called Reggeon field theory for classical DP \cite{Reggeon}. It features --- upon rescaling the fields --- the characteristic \emph{rapidity inversion} symmetry, which leaves the system invariant under the transformation $n\leftrightarrow-\tilde{n}$ and $t\rightarrow-t$ % which replaces microreversibility and detailed-balance for DP-like processes
 \cite{Tauber-book, Kamenev, Sieberer15}. For $u_4>0$, this symmetry is broken by the microscopic action.
The implications depend on the dimension $d$: For $d>2$, $u_4$ is RG irrelevant and can be discarded in the infrared-dominated dynamics close to the second order transition. Consequently, in $d>2$, rapidity-inversion is restored and the line of continuous transitions displays universal scaling behavior corresponding to classical DP.

At the $\alpha$ point [white dot in Fig.~\ref{fig:1}(b)], $u_3=0$ and the leading-order coupling becomes $u_4$. For $d>2$, the second order transition at this point is governed by mean-field scaling behavior, since $u_4$ is RG-irrelevant and cannot introduce infrared divergent corrections to the vanishing couplings $u_3, \Delta$.
On the other hand, for $d<2$, $u_4$ becomes RG relevant and generates a non-trivial RG flow of $\Delta$ and $u_3$ on the entire second order transition line.
This leads to a violation of rapidity-inversion which persists at long wavelength, and thus drives the system away from the DP critical point to a new non-equilibrium universality class, without specific symmetries. In $d<2$, therefore, only the isolated point $\chi=1/\nn$, $\omega=0$ lies in the DP class, while the presence of quantum fluctuations imprints a new universal scaling behavior on the entire line, including the $\alpha$ point. For marginal $u_4$ in $d=2$, the scaling of the fluctuation corrections to $u_4$ determines whether this coupling becomes relevant, making the scenario equivalent to $d<2$, or irrelevant, which has to be determined by an RG analysis of the problem.

\section{Non-equilibrium discontinuous transition}
For $(\Delta > 0, u_3 < -2\sqrt{\Delta u_4})$ the effective potential $\Gamma$ displays two distinct minima, $n_1=0$ and $n_2=\tfrac{|u_3|}{2u_4}+(\tfrac{u^2_3}{4u_4^2}-\tfrac{\Delta}{u_4})^{1/2}$, suggesting a first-order phase transition. 
%The thin, dashed line in Fig.~\ref{fig:1}(b) corresponds to the appearance of the second minimum. 
The actual transition line, however, lies where the finite-density minimum becomes statistically preferred. In equilibrium, this would be the point at which the minima of $\Gamma$ are at the same height.
However, the present non-equilibrium noise shows more pronounced fluctuations at larger densities and thus favors the absorbing minimum $n_1$ with respect to $n_2$. To estimate the steady state distribution function $P(n)$, we apply the optimal path approximation to the action \cite{Kamenev,Tauber-book}; this involves treating the coefficient $\Xi(n) = \tfrac{1}{2} n + \mu_4 n^2$ of $\tilde{n}^2$ as a kind of mean-field and density-dependent temperature.
It yields (see Appendix \ref{app:opt_path})
\begin{eqnarray}
P(n)=\tfrac{1}{Z}e^{-V\,W(n)},\,\,\, \text{with}\,\,\, W(n)=\int_0^ndm \tfrac{\partial\Gamma/ \partial m}{\Xi(m)},\label{opath}
\end{eqnarray}
with volume $V$ and normalization $Z$. Both potentials $W(n)$ and $\Gamma(n)$ vanish in $n_1$ and share the finite-density minimum $n_2$. In the thermodynamic limit $V\rightarrow\infty$, $P(n)\rightarrow\delta(n-n_l)$, where $l=1,2$ depending on which one is the global minimum of $W$, accounting for the physical constraint $n \geq 0$. The transition takes place when $W(n_2) = 0$, which identifies the non-equilibrium first-order line [dashed line in Fig.~\ref{fig:1}(b)]. Due to the non-equilibrium nature of the fluctuations, this does not coincide with the naive prediction $\Gamma(n_2) = 0$, as shown in Fig.~\ref{fig:2}(c). In Fig.~\ref{fig:2}(d) we report the full-counting statistics of the density $n$ obtained via QJMC techniques \cite{Q_jump} for a chain of $12$ spins. Despite the presence of strong finite-size effects, a bimodal structure is still highlighted for large values of $\omega$. This implies that trajectories bunch together around two possible values, the absorbing one and a finite-density one and is a signature of the aforementioned coexistence.

\section{Realization with Rydberg atoms} 
Instances of this crucial competition between classical and quantum processes can be implemented with cold atoms excited to Rydberg states \cite{Viteau11, PRL-KinC, Lesanovsky14, Olmos14-1, Marcuzzi14, Schaub2015}. They are represented with two internal states, the ground state $\ket{\text{GS}} \equiv \ket{i}$ (inactive site) and the excited one $\ket{\text{Ryd}} \equiv \ket{a}$ (active site). Rydberg gases feature strong van-der-Waals interactions in state $\left|a\right>$ \cite{Gallagher84, Rydberg2, Ryd-QI}, which rapidly decay as $r^{-6}$ with the interparticle distance $r$. For the sake of simplicity, we approximate it here as a nearest-neighbor interaction of strength $V_\mathrm{nn}$ in a one-dimensional configuration.

Quantum branching/coagulation is realized via coherent driving by a laser field of Rabi frequency $\Omega$ and detuning $\Delta_L$ with respect to the atomic transition frequency; fixing $\Delta_L  = -V_\mathrm{nn}$ enables an ``anti-blockade'' \cite{Ates07, Amthor2010, Lesanovsky14} mechanism which favors the excitation of a Rydberg atom next to an already excited one, e.g. $\left|iai\right>\rightarrow \left|iaa\right>$. Differently from the idealized model above, the constraint requires here a \emph{single} excitation nearby, and processes such as $\left|aia\right>\rightarrow \left|aaa\right>$ are highly suppressed. The Hamiltonian is therefore approximately given by $H_\mathrm{ryd}=\Omega \sum_k \Pi'_k \sigma^x_k $ where $\Pi'_k=n_{k-1}+n_{k+1}-2n_{k-1}n_{k+1}$.

To generate the incoherent branching/coagulation the atoms are coupled (with coupling $g$) to a second equally-detuned light field with strong phase noise (dephasing rate $\lambda \gg g$) \cite{Walls85}; for a correlation length shorter than the interatomic distance, the bath is modeled as independent bosonic modes $b_k$, $b_k^\dag$ acting on each lattice site.
The effective equation of motion for the atoms is obtained by performing second order perturbation theory in the small parameter $g/\lambda$ \cite{Degenfeld14, Marcuzzi14, Marcuzzi2015}. 
The resulting master equation for the reduced atomic density matrix $\rho$ is
\begin{eqnarray*}
  \dot{\rho} &=& \frac{4 g^2}{\lambda}\sum_k \left( \langle b^\dagger_k b_k \rangle\mathbf{D}[\Pi'_k \sigma^+_k]+\langle b^\dagger_k b_k+1\rangle\mathbf{D}[\Pi'_k \sigma^-_k]\right) \rho.
\end{eqnarray*}
For sufficiently high ($\langle b^\dagger_k b_k \rangle \gg 1$) and homogeneous ($\langle b^\dagger_k b_k \rangle \approx \langle b^\dagger_m b_m \rangle$) intensity, one can identify $\kappa=(4 g^2 \langle b^\dagger_k b_k \rangle)/\lambda$, leading to the branching/coagulation jump operators: $L^\mathrm{ryd}_{\text{b},k} = \sqrt{\kappa} \, \Pi'_k \, \sigma_k^{+}$ and $L^\mathrm{ryd}_{\text{c},k} = \sqrt{\kappa} \, \Pi'_k \, \sigma_k^{-}$.
The final process is radiative decay of an atom from its Rydberg state to the ground state, modeled by the jump operator $L^\mathrm{ryd}_{\text{d},k} = \sqrt{\gamma} \, \sigma_k^{-}$ \cite{Rydberg2}.

Although the microscopic formulation of the dynamics is slightly different from the previously-discussed model, the resulting phase structure is similar, as the EOMs only differ from Eqs.~(\ref{eq:quantum_n}-\ref{eq:quantum_y}) by RG irrelevant higher order density terms. In particular, they leave the universal properties near the continuous transition points unchanged.

\section{Outlook} We have investigated the effects of quantum dynamical processes on a prototypical absorbing-state phase transition. We highlighted the emergence of a richer structure in the phase diagram, which includes both a discontinuous and a continuous non-equilibrium transition. In low dimension $d<2$ the presence of a quantum coherent process leads to a breaking of the only fundamental symmetry of DP in a way that persists at long wavelengths, and thus leads to a phase transition of  a different nature. In equilibrium, the interplay between classical (thermal) and quantum fluctuations typically leads to a dimensional crossover \cite{QPT, Chiocchetta2015}. The present work shows that out of equilibrium the picture is not as straightforward and opens the path for further investigations in this field, including the quantitative characterization of the new universality class.

\acknowledgments
M.M.~and I.L.~wish to express their gratitude for the insightful discussions with J.P.~Garrahan and for access to the University of Nottingham High Performance Computing Facility. I.L.~acknowledges that the research leading to these results has received funding from the European Research Council under the European Union's Seventh Framework Programme (FP/2007-2013) / ERC Grant Agreement n. 335266 (ESCQUMA). Further funding was received through the H2020-FETPROACT-2014 grant No.~640378 (RYSQ) and from EPSRC Grant no.\ EP/J009776/1. M. B. and S. D. acknowledge funding by the German Research Foundation (DFG) through the Institutional Strategy of the University of Cologne within the German Excellence Initiative (ZUK 81).

%%%%%%%%%%%%%%%%%%%%%%%%%%%%%%%%%%%%%%%%%%%%%%%%%%%%%%%%%%%%%%%%%%%%%%%%%%%%%%%%%%%%%%%%%%%%%%%%%%%

%%%%%%%%%%%%%%%%%%%%%%%%%%%%%%%%%%%%%%%%%%%%%%%%%%%%%%%%%%%%%%%%%%%%%%%%%%%%%%%%%%%%%%%%%%%%%%%%%%%

%%%%%%%%%%%%%%%%%%%%%%%%%%%%%%%%%%%%%%%%%%%%%%%%%%%%%%%%%%%%%%%%%%%%%%%%%%%%%%%%%%%%%%%%%%%%%%%%%%%

%%%%%%%%%%%%%%%%%%%%%%%%%%%%%%%%%%%%%%%%%%%%%%%%%%%%%%%%%%%%%%%%%%%%%%%%%%%%%%%%%%%%%%%%%%%%%%%%%%%

%%%%%%%%%%%%%%%%%%%%%%%%%%%%%%%%%%%%%%%%%%%%%%%%%%%%%%%%%%%%%%%%%%%%%%%%%%%%%%%%%%%%%%%%%%%%%%%%%%%

%%%%%%%%%%%%%%%%%%%%%%%%%%%%%%%%%%%%%%%%%%%%%%%%%%%%%%%%%%%%%%%%%%%%%%%%%%%%%%%%%%%%%%%%%%%%%%%%%%%

%%%%%%%%%%%%%%%%%%%%%%%%%%%%%%%%%%%%%%%%%%%%%%%%%%%%%%%%%%%%%%%%%%%%%%%%%%%%%%%%%%%%%%%%%%%%%%%%%%%

%%%%%%%%%%%%%%%%%%%%%%%%%%%%%%%%%%%%%%%%%%%%%%%%%%%%%%%%%%%%%%%%%%%%%%%%%%%%%%%%%%%%%%%%%%%%%%%%%%%

\begin{appendices}

\begin{section}{Heisenberg-Langevin Equations of Motion}
\label{app:HL_eqs}

In order to derive the Heisenberg-Langevin equations of motion of an observable $O$ we employ the conjugate Master equation
\begin{equation}
	\dot{O} = i \comm{H}{ O} + \suml{a,k}{} \mathbf{D}'[L_{a,k}]O+\hat{\xi}^O,
\end{equation}
where $\mathbf{D}'[X] O =  X^\dagger O X - (X^\dagger X O + O X^\dagger X)/2$ and $\hat{\xi}^O$ is the quantum noise term for the operator $O$. The noise-less equation of motion is only formally correct on the level of single-operator expectation values, while the noise contributes by preserving the (anti-)commutation relations of the operators during the evolution.
%takes into account that generally $O(t)O(t)\neq O^2 (t)$ and implements corrections for higher order expectation values 
\cite{Scully}. For a linear coupling of the system to the bath, the noise is typically Gaussian, with zero mean, but non-vanishing time- and space-local correlations. 
 This prescription leads to Eqs.~\eqref{eq:quantum_n}-\eqref{eq:quantum_y}; below we provide the main conceptual steps.

There exist several canonical (and equivalent) strategies to determine the properties of the noise operators $\hat{\xi}^{n,x,y}$, as for instance outlined in Ref.~\cite{Scully}. Here, we follow a path relying on the unitary Heisenberg equations of motion for system operators in the presence of a bath. As a simplifying assumption, we imagine the spatial correlations of this bath to be shorter than the typical interparticle distance in the system. This allows us to describe every spin as coupled to its own bath. We can therefore focus on a single spin as a representative and we
model the spontaneous emission dynamics via the simple Hamiltonian
\begin{eqnarray}
\widetilde{H}=H_{\text{s-b}}+H_{\text{b}}=\sum_q \lambda_q(\sigma^+b_q+b^{\dagger}_q\sigma^-)+\sum_q \omega_q b^{\dagger}_q b_q, \ \ \ \ \ \ 
\end{eqnarray}
where the $b_l$s represent a set of bosonic bath operators. We further assume that this bosonic reservoir is kept at zero temperature and that the number of modes is sufficiently large to allow a continuum description with a density of states $D(\omega) = \sum_q \delta (\omega - \omega_q)$. Taking the von Neumann equation for the global (spin plus bath) density matrix and eliminating the bath degrees of freedom leads then to the jump operator $L_d=\sqrt{\gamma}\sigma^-$ with $\gamma=2\pi [\lambda(0)]^2D(0)$ being proportional to the bath density of states $D(0)$ and the couplings $\lambda(0)$ evaluated at zero frequency (see e.g.~Chapter $8$ of \cite{Scully}).
The Heisenberg equations of motion for the operators are therefore
\begin{eqnarray}
\dot{\sigma}^+&=&i[\widetilde{H},\sigma^+]  =   - i\sum_q \lambda_q  b^{\dagger}_q \sigma^z  \label{sigp},\\
\dot{n}&=&i\sum_q\lambda_q (b^{\dagger}_q \sigma^-  - \sigma^+ b_q),\label{dens}\\
\dot{b}^{\dagger}_q&=&i\lambda_q\sigma^+ + i\omega_q b^{\dagger}_q.\label{DaggerEq}
\end{eqnarray}
Formally solving Eq.~\eqref{DaggerEq} yields
\begin{eqnarray}
b^{\dagger}_q(t) =  b_q^{\dagger}(0)  e^{i\omega_q t}+i\lambda_q \int_0^t dt' \sigma^+(t')e^{i\omega_q(t-t')}.
\end{eqnarray}
Inserting this solution into Eqs.~\eqref{sigp}, \eqref{dens} and performing the Born-Markov approximation leads to
\begin{align}
\dot{\sigma}^+& = -\frac{\gamma}{2}\sigma^{+}+\underbrace{i\sum_q \lambda_q  b^{\dagger}_q (0) \sigma^ze^{i\omega_q t}}_{\tilde{\xi}^+(t)},\\
\dot{n}& = -\gamma n +\underbrace{i\sum_q \lambda_q (b^{\dagger}_q (0) \sigma^-e^{i\omega_q t}-\sigma^+b_q (0) e^{-i\omega_q t})}_{\tilde{\xi}^n(t)}.
\end{align}
Defining $\tilde{\xi}^{-}(t)=(\tilde{\xi}^+(t))^{\dagger}$ and taking the bath to be in the vacuum state (corresponding to spontaneous emission), we find therefore the noise properties in the Born Markov approximation to be
\begin{eqnarray}
\langle\tilde{\xi}^+(t)\tilde{\xi}^+(t')\rangle&=&\langle\tilde{\xi}^+(t)\tilde{\xi}^-(t')\rangle=\langle \tilde{\xi}^+(t)\rangle=\langle \tilde{\xi}^n(t)\rangle=0,\nonumber\\
\langle\tilde{\xi}^-(t')\tilde{\xi}^{+}(t)\rangle&=&\gamma\delta(t-t').
\end{eqnarray}
By noticing that $\tilde{\xi}^n \equiv -\tilde{\xi}^+ \sigma^- - \sigma^+ \tilde{\xi}^-$ one gets all the remaining non-vanishing correlations
\begin{eqnarray}
\langle\tilde{\xi}^n(t)\tilde{\xi}^n(t')\rangle&=&\gamma n \delta(t-t'), \label{eq:n_corr}\\
\langle\tilde{\xi}^n(t)\tilde{\xi}^+(t')\rangle&=& - \gamma \sigma^+ \delta(t-t'), \\ 
\langle\tilde{\xi}^n(t)\tilde{\xi}^+(t')\rangle&=& - \gamma \sigma^- \delta(t-t').
\end{eqnarray}
Rotating into the $(x,y,n)$ basis and introducing $\tilde{\xi}^x=\tilde{\xi}^++\tilde{\xi}^-$ and $\tilde{\xi}^y = - i\tilde{\xi}^+  + i\tilde{\xi}^-$ (and analogously $\sigma^x$ and $\sigma^y$) one finds $\av{\tilde{\xi}^i(t) \tilde{\xi}^j(t')} = \gamma \delta (t-t') M^{ij}$ with
\begin{equation}
	M = \matb{ccc} 1 & -i & -\frac{\sigma^x - i \sigma^y}{2} \\ i & 1 & -\frac{i\sigma^x + \sigma^y}{2} \\ - \frac{\sigma^x + i\sigma^y}{2} & i\frac{\sigma^x + i\sigma^y}{2} & n    \mate.
\end{equation}
The dependence of the $\xi^n$ noise on the density keeps track of the fact that the absorbing configuration $n=0$ represents a fluctuationless state in the entire parameter regime, which forbids a density independent contribution to $\xi^n$ in Eq.~\eqref{eq:n_corr}. The Markovian noise level introduced by the decay terms $L_d=\gamma \sigma^-$ only appears as an additive noise in the $\sigma^{x,y}$ variables.

Including classical coagulation and branching processes yields additional noise terms. However, due to the presence of the term $\sum_j n_j$ these contributions will always be higher-order in the density and are therefore subleading with respect to the ones derived above in the absorbing phase. Extending the system from a single spin to a lattice of individual spins, an equivalent computation shows 
\begin{eqnarray}
&\langle\hat{\xi}^n_k (t) \hat{\xi}^n_{k'} (t') \rangle = n_k \gamma \delta_{k,k'} \delta (t-t') + O(n^2),\\
&\langle\hat{\xi}^x_k\hat{\xi}^x_{k'}\rangle  =  \gamma \delta_{k,k'} \delta (t-t') + O(n) = \langle\hat{\xi}^y_k\hat{\xi}^y_{k'}\rangle.
\end{eqnarray}
To leading order in the density, this yields the same noise terms reported above. Since the coherent branching and coagulation does not produce an additional noise, this concludes the derivation of the Heisenberg-Langevin equations.

\end{section}

\begin{section}{Martin-Siggia-Rose construction}
\label{app:path_int}

In this section, we provide the derivation of the Martin-Siggia-Rose (MSR) path integral for the present quantum contact process, which results in the effective long wavelength action for the density, Eq.~\eqref{eq:action}. As a first step, we take the continuum limit of the equations of motion for $n_k$, $\sigma^x_k$ and $\sigma^y_k$, such that
\begin{eqnarray}
\sum_{j \text{ nn } x} n_j \to  (r^2\nabla^2  +  z)n_x,
\end{eqnarray}
where $z$ is the coordination number, $r$ is the lattice spacing, $\nabla$ is the the common $d$-dimensional gradient and $x = rk$ the position. We then re-interpret the operators as stochastic fields subject to the continuum noise sources $\xi^x_X$, $\xi^y_X$, $\xi^n_X$ --- where $X=(t,x)$ is shorthand for the spatio-temporal argument --- which have vanishing mean and correlations $\av{\xi^i_X \xi^j_Y} = \gamma \delta (X - Y) M^{ij}$, where
\begin{equation}
	M =   \matb{ccc} 1 & 0 & -\frac{\sigma^x }{2} \\ 0 & 1 & -\frac{ \sigma^y}{2} \\ - \frac{\sigma^x }{2} & -\frac{\sigma^y}{2} & n    \mate.
\end{equation}
The equations of motion can thus be expressed as
\begin{align}
\dot{n}_X& = \mathcal{F}_n(n_{X},\sigma^x_X , \sigma^y_{X}) + \xi^n_X,  \label{eq:Sn}\\
\dot{\sigma}^x_X&= \mathcal{F}_{\sigma^x}(n_{X},\sigma^x_X , \sigma^y_{X}) + \xi^x_X,   \label{eq:Sx}\\
\dot{\sigma}^y_X&= \mathcal{F}_{\sigma^y}(n_{X},\sigma^x_X , \sigma^y_{X}) + \xi^y_X,  \label{eq:Sy}
\end{align}
where
\begin{align}
 &\mathcal{F}_n =  -n_X+[\omega\sigma^y_X-\chi(2n_X-1)](r^2\nabla^2+z)n_X, \\[2mm]
&\mathcal{F}_{\sigma^x} = -\tfrac{z\chi+1}{2}\sigma^x_X  -  \chi\sigma^x_X (r^2\nabla^2+z)n_X  + \nol 
& \ \ \ + \omega \sigma^y_X (r^2 \nabla^2 + z) \sigma^x_X, \\[2mm]
 &\mathcal{F}_{\sigma^y} = -\tfrac{z\chi+1}{2}\sigma^y_X-[\omega(4n_X  -   2)+\chi\sigma^y_X](r^2\nabla^2+z)n_X + \nol 
 & \ \ \  + \omega \sigma^x_X (r^2 \nabla^2 + z) \sigma^x_X.
\end{align}

As shown in Refs.~\cite{Kamenev, Tauber-book}, the MSR construction defines a path integral in the variables $\sigma^y_X, n_X$ for the equations of motion \eqref{eq:Sn}-\eqref{eq:Sy}. The MSR partition function represents the sum over all allowed field configurations, i.e.
\begin{align}
&Z  =   \int \mathcal{D}[n_{X}, \sigma^x_X , \sigma^y_{X}]  \int\mathcal{D}[\xi^n_X , \xi^x_X , \xi^y_X] P(\xi^n_X , \xi^x_X , \xi^y_X) \times  \nonumber \\
 & \times J[n_{X}, \sigma^x_X , \sigma^y_{X}]  \, \delta\left( \dot n_{X}-\mathcal{F}_n(n_{X}, \sigma^x_{X} , \sigma^y_{X})\right) \times \nol[2mm]
 &\times   \delta\left(\dot \sigma^x_{X}-\mathcal{F}_{\sigma^x}(n_{X}, \sigma^x_{X} , \sigma^y_{X})\right) \, \delta\left(\dot \sigma^y_{X}-\mathcal{F}_{\sigma^y} (n_{X}, \sigma^x_{X} , \sigma^y_{X})\right),
\end{align}
where the integral $\int\mathcal{D}[\vec{\xi}]P(\vec{\xi})$ averages over all noise configurations described by the Gaussian noise distribution $P(\vec{\xi}) = \exp  \left\{ - \tfrac{1}{2} \int_X  \lt \vec{\xi}_X \rt^\intercal M^{-1} \vec{\xi}_X    \right\} $. The factor $J[n_{X},\sigma^y_{X}]$ is a Jacobian which, for our present purposes, can be conveniently set to $1$ after choosing a proper, retarded regularization \cite{Kamenev, Tauber-book}.
Introducing three sets of imaginary response fields $\tilde{n}_{X}, \tilde{\sigma}^x_X, \tilde{\sigma}^y_{X}$, exploiting the Fourier transform $\delta(f(n)) = \int \mathcal{D} \tilde{n} \exp(-\tilde{n} f(n))$ and integrating over the noise variables $\xi^{n/x/y}$, $Z$ can be cast into a path-integral form
\begin{eqnarray}
\label{eq:Z}
Z= \int \mathcal{D}[\tilde{n}_{X},n_{X},\tilde{\sigma}^x_{X}\sigma^x_{X},\tilde{\sigma}^y_{X}\sigma^y_{X}]\ e^{-S}.
\end{eqnarray}
where, up to leading order in a (spatial) derivative expansion, the action reads
\begin{widetext}
\begin{equation}
	\begin{split}
	 S &= \int_X \tilde{n}_X \left[  \partial_t  - D\nabla^2  - (z\chi -1)-  \ha\tilde{n_X}   \right] n_X   + 2z\chi \tilde{n}_X n_X^2  \\[2mm]
	& + \int_X \tilde{\sigma}^y_X \lqq   \partial_t + \tfrac{z\chi+1}{2} + z\chi n_X    +  \ha  \tilde{n}_X \rqq \sigma^y_X -\ha   \lt \tilde{\sigma}^y_X \rt^2   +    \tilde{\sigma}^y_X (z \omega (4n^2_X - 2n_X)) - \sigma^y_X (z\omega \tilde{n}_X n_X)    \\[2mm]
	& +  \int_X    \tilde{\sigma}^x_X  \lqq  \partial_t + \tfrac{z\chi+1}{2} + z\chi n_X   + z\omega \sigma^y_X   +  \ha  \tilde{n}_X \rqq \sigma^x_X   -\ha   \lt \tilde{\sigma}^x_X \rt^2      - z\omega  \tilde{\sigma}^y_X \lt\sigma^x_X  \rt^2  .    
\end{split}
\end{equation}
\end{widetext}
Since $\tfrac{z\chi + 1}{2} \geq 1/2$ throughout the physical parameter region $\chi \geq 0$, both the $\sigma^x$ and the $\sigma^y$ fields remain gapped and one can therefore neglect the subleading derivative and fluctuating terms within square brackets. This yields an action which is separately quadratic in $\lt \sigma^x_X, \tilde{\sigma}^x_X  \rt$ and $\lt \sigma^y_X, \tilde{\sigma}^y_X  \rt$. These modes can be actually integrated out exactly and, up to RG-irrelevant terms, one obtains action \eqref{eq:action}. The couplings correspond to the following combinations of miscroscopic parameters:
\begin{subequations}
\begin{align}
	\Delta  &= 1 - z\chi - \tfrac{8z^2\omega^2}{(z\chi+1)^3}, \\
	u_3 &= 2z \left(\chi- \tfrac{2z\omega^2}{z\chi+1}\right),  \\
	 u_4 &= \tfrac{ 8z^2 \omega^2}{z\chi+1}, \\
	 \mu_4 & =  \frac{2z^2 \omega^2}{(z\chi + 1)^2} +   \frac{128 z^4 \omega^4}{(z\chi + 1)^6}.
\end{align}
\end{subequations}
Our procedure differs conceptually from the approach advocated in \cite{maghrebi2015}. We found it necessary to accurately capture the short distance physics of the problem.

\end{section}

\begin{section}{Additional details on the nature of the observed phase transitions}
\label{app:add_details}

Except for the $\alpha$ point, in the proximity of the transitions $u_3 \neq 0$ and one can rescale the fields according to $n\rightarrow K n$, $\tilde{n}\rightarrow\tilde{n}/K$ with the choice $K=1/\sqrt{2u_3}$. Thus, one finds
\begin{eqnarray}\label{Eq23a}
S&=&\int_X \hspace{-0.2cm}\tilde{n}_X\left[\partial_t-D\nabla^2+\Delta+\kappa_3n_X+\tfrac{u_4}{2u_3}n_X^2\right]n_X \nonumber\\
&&-\int_{X}\hspace{-0.1cm}\tilde{n}_X^2 \left[\kappa_3n_X+ \mu_4 n_X^2 \right].\ \ \
\end{eqnarray}
with $\kappa_3 = K u_3 =1 / 2K = \sqrt{u_3/2}$.
In the absence of the $u_4$ coupling, i.e. for $\omega=0$, this is the action describing classical directed percolation.
It features the characteristic rapidity inversion symmetry under $n\leftrightarrow-\tilde{n}$, $t\rightarrow-t$.
%As a consequence, for $\omega=0$, the system falls into the universality class of classical directed percolation.
For $\omega>0$, the relevance of the $u_4$ coupling has to be considered, which is determined by the scaling behavior of the fields $n, \tilde{n}$. In the absence of a thermal fluctuation dissipation relation, both fields $n, \tilde{n}$ typically have the same scaling dimension \cite{Kamenev, Tauber-book, Altland}.
This leads to an upper critical dimension of $d_c=4$ for the cubic coupling $u_3$ and an upper critical dimension  of $d_c=2$ for the quartic coupling $u_4$. In dimensions $d>2$, $u_4$ renormalizes to zero in the RG flow and the rapidity inversion symmetry is restored in the infrared regime. Hence, the effective low frequency theory, and therefore the long time dynamics, is again described by the directed percolation class. On the other hand, for $d<2$, $u_4$ is relevant in the renormalization group sense and the absence of rapidity inversion introduces a different non-equilibrium dynamics at the phase transition, which is not captured by the DP universality class. In $d=2$, the quartic couplings are marginal and whether they become relevant or irrelevant in the RG flow has to be determined by a renormalization group analysis of the problem.

At the point $\alpha$, $u_3$ vanishes microscopically and the rescaling leading to the action \eqref{Eq23a} is not defined. In dimensions $d>2$ this point features a second order phase transition in the absence of the rapidity inversion symmetry. Since the leading order term in the effective potential $\Gamma$ (Eq.~\eqref{Eq10}) is RG irrelevant in $d>2$, one expects mean-field scaling behavior at this point. On the other hand, in $d<2$ the effective theory at the $\alpha$-point corresponds to the new universality class in the absence of rapidity inversion.

In any experiments with cold atoms, the presence of small fluctuation-inducing terms $\sim \Delta_x\sigma^x$ or $\Delta_y\sigma^y$ is hardly avoidable at the microscopic level. These will generate fluctuations on top of the absorbing state and lead to a temperature type term $\sim \tilde{n}_X^2 T$ in the action \eqref{Eq23a}, with $T\approx\Delta_x,\Delta_y$. The present discussion of the non-equilibrium phase transitions is then valid on length scales $l^{-1}\ge\sqrt{T}$.

\end{section}

\begin{section}{Details on optimal path approximation}
\label{app:opt_path}
In the present setting, the noise $\Xi_X\equiv  \tfrac{1}{2} n_X+ \mu_4 n_X^2$ increases monotonically with the density. As a consequence, it favors the (fluctuationless) zero density solution over the finite-density one. In order to determine the distribution function for the density variable in the vicinity of the active-to-inactive transition, we apply the optimal path approximation \cite{Kamenev, Tauber-book} to the partition function. Note that the system remains gapped for $\Delta > 0$ (where the first-order transition is expected to take place) and therefore we can --- as a first approximation --- neglect spatial fluctuations and approximate $n_X, \tilde{n}_X$ by spatially homogeneous but temporally fluctuating fields $n_t, \tilde{n}_t$. This yields the action
\begin{eqnarray}
S=V\int_t \Big[\tilde{n}_t\partial_t n_t+\tilde{n}_t\Gamma'(n_t)-\tilde{n}_t^2\Xi(n_t)\Big]
\end{eqnarray}
with the shorthand $\Gamma'(n) = \delta \Gamma / \delta n$.
The optimal path for the configurations $n_t, \tilde{n}_t$ corresponds to the configurations for which the non-fluctuating part of the action vanishes, i.e. for which ${\tilde{n}_t\Gamma'(n_t)-\tilde{n}_t^2\Xi(n_t)=0}$.
This equation shows the two trivial solutions $\tilde{n}_t=0$ and $n_t$ arbitrary as well as $n_t=0$ and $\tilde{n}_t$ arbitrary. Apart from this, there exists the non-trivial solution
\begin{eqnarray}
\tilde{n}_t^{\text{op}}=\frac{\Gamma'(n_t)}{\Xi(n_t)}.
\end{eqnarray}
Considering only configurations which correspond to the optimal path the action becomes
\begin{eqnarray}
W(n)=\int_0^\infty \tilde{n}_t^{op}\partial_t n_t \rmd t = \int_{n_0}^{n} \frac{\Gamma'(m)}{\Xi(m)} \rmd m,
\end{eqnarray}
with the change of variable $\partial_t n_t \rmd t \to \rmd m$ and where $n_0$ is the initial condition --- whose specific value is irrelevant --- and $n$ the steady-state value of the density.
The corresponding density distribution function is
\begin{eqnarray}
P(n)=\frac{1}{Z}  e^{-VW(n)},
\end{eqnarray}
with $Z=\int_ne^{-VW(n)}$, if the integral exists. For a system in thermal equilibrium, $\Xi(n) \propto T$ is simply proportional to the temperature and one recovers the naive expectation $P(n) \sim \exp (-\Gamma(n)/T)$.
For the present noise terms
\begin{eqnarray}
W(n) = \tfrac{1}{\mu_4}\Big[\Delta l+u_3(n - \tfrac{l}{2\mu_4})+u_4 \tfrac{ 4n \mu_4 (n\mu_4 - 1) + 2l}{16 \mu_4^3}
\Big],\ \ \ \ \ \ \
\end{eqnarray}
with $l=\log[1+ 2n \mu_4]$.

The minima of $W(n)$ are $n_1 = 0 $ and $n_2 = -\tfrac{u_3}{2u_4}+\sqrt{\tfrac{u_3^2}{4u_4^2}-\tfrac{\Delta}{u_4}}$ and coincide with the ones of $\Gamma(n)$. The two functionals, however, may differ significantly. In particular, the global minimum of $W$ does not necessarily coincide with the global minimum of $\Gamma$, such that the presence of a non-equilibrium noise term can strongly modify the phase boundary as a function of the noise strength. Since $W(0)=0$ for all parameters, the first order transition line separating the active from the inactive phase for $\Delta > 0$ is determined by the equation $W(n_2)=0$.

\end{section}

\end{appendices}

\bibliography{biblio}

\end{document}